\documentclass[11pt,twoside]{article}
\usepackage{asp2010}

\resetcounters

\begin{document}

\title{The violent neutron star}
\author{Anna L. Watts,$^1$
\affil{$^1$Astronomical Institute 'Anton Pannekoek', University of Amsterdam, Science Park 904, 1090GE Amsterdam, The Netherlands.  Email: A.L.Watts@uva.nl}}

\begin{abstract}
Neutron stars enable us to study both the highest densities and the highest magnetic fields in the known Universe.  In this article I review what can be learned about such fundamental physics using magnetar bursts.  Both the instability mechanisms that trigger the bursts, and the subsequent dynamical and radiative response of the star, can be used to explore stellar and magnetospheric structure and composition.  
\end{abstract}

\section{The neutron star laboratory}

Neutron stars have densities so high that nuclei, and even neutrons, may dissolve to form exotic
states of matter. The high densities also allow neutron stars to sustain magnetic fields ten orders
of magnitude higher than those we can create in terrestrial facilities, in regimes where unusual
electromagnetic processes are expected.  These properties allow us to explore physics that cannot be studied in the laboratory.  

\subsection{Dense matter}

The theory of Quantum Chromodynamics provides an excellent description of how quarks are bound together to form nucleons.  Understanding the interaction between nucleons, however, is much more complex.  Many-body nucleon theory (even at the level of the three-nucleon interaction), particularly for isospin asymmetric matter and for densities exceeding the saturation density, is especially challenging.  The nature of this interaction, and the state of matter at extremes of temperature and density, are extremely active fields of research whose resolution requires an integrated effort between collider experiments and relativistic astrophysics.   Of particular interest is the possibility of transitions from nucleons to de-confined quarks and gluons or other more exotic states. At low densities this can be studied by experiments like the Large Hadron Collider or in various Heavy Ion experiments such as the Relativistic Heavy Ion Collider and the Facility for Antiproton and Ion Research.  At higher densities, however, neutron stars are the only environment in the Universe where the transition can be explored.   

\subsection{Strong magnetic fields}

The strongest magnetic fields that can be studied within our Solar System are those
created in high magnetic field laboratories on Earth. At present these facilities can generate
fields approaching $10^6$ G, but only for fractions of a second. Neutron stars are the only objects
that let us study fields above $10^9$ G, and are the only stars with fields exceeding the quantum critical limit of
$B_\mathrm{QED} = 4.4\times10^{13}$ G.  At this point the magnetic field energy is so high that it can generate electron-positron
pairs, so that the vacuum seethes with charged particles.  In such an environment we
expect a host of unusual and intriguing physical effects such as spontaneous photon splitting and vacuum birefringence.  

\section{The power of violent dynamical events}

There are many different ways of using neutron stars to study both dense matter and strong magnetic fields.    One method is to harness the power of violent dynamical events that manifest as bursts or flares.  Such events involve instabilities in physical processes that are often extremely interesting in their own right and may depend crucially on aspects of the physics that we want to study.  We can also however analyse the subsequent dynamical and radiative response of the star to obtain even deeper insight.   

In this review I will focus exclusively on bursts from magnetars, isolated neutron stars with the very highest magnetic fields, often exceeding $B_\mathrm{QED}$ \citep{Woods06, Mereghetti08}.  Magnetars are
highly active, with spectacular outbursts of gamma-ray bursts powered by decay of the magnetic
field.  These bursts can be so violent that they excite long-lived seismic vibrations of the star and its magnetosphere \citep{Israel05, Strohmayer05, Strohmayer06, Watts06}.  

\subsection{Instabilities and the magnetar burst trigger}

The underlying cause of magnetar activity is decay of the ultra-strong magnetic field, which twists
the field lines into an unstable configuration \citep{Braithwaite06}. Once a tipping point is
reached, the field lines undergo rapid evolution, reconfiguration and possibly reconnection.  This creates currents whose dissipation generates gamma-rays. What is not
understood, however, is what triggers the bursts. For flaring to be sporadic, there has to be
some barrier to magnetic reconfiguration that yields when a threshold is reached \citep{Duncan04}.

The solid crust
can in principle resist motion as it is stressed by the changing interior field, yielding only when
magnetic force exceeds the breaking strain \citep{Thompson95}. The resulting crust
rupture enables external field lines to move and reconfigure, generating the flare. Whether this is
the case depends on the breaking strain of the crust, set by its composition, crystalline structure, and
melting properties \citep{Horowitz09}. The strength of the deep crust, where the nuclei are
expected to be highly deformed (the so-called pasta phase), is likely to be critical \citep{Pethick98}.

Stress could also build up in the external magnetosphere,
being released only when plasma conditions permit reconnection via various instabilities \citep{Lyutikov03, Gill10, Levin12}. If the crust yields plastically rather than resisting stress, then this may be more likely \citep{Jones03}.  The stress release mechanism must also permit events of different sizes, from the very shortest weakest bursts up to the most energetic giant flares.    Bursts can occur singly or sometimes in storms, where multiple events occur very close together \citep{Israel08, Kaneko10}.     It is also clear that bursting is not the only way of transferring magnetic stress 
from the interior of the star to the exterior. The level of twist in the magnetosphere changes even
when there is no flaring, indicating that non-violent stress transfer is also possible \citep{Thompson02}.

Pinpointing the mechanism that triggers the bursts has been hampered by a lack of quantitative
predictions for flare lightcurves and spectra. This is due in part to the complex effects that fields above the quantum 
critical value have on radiative transfer \citep{Baring95, Heyl05, Harding06}.  The magnetic energy is high enough for the spontaneous creation of electron/positron pairs,
and charged particles are constrained to move only along magnetic field lines. Photons interacting
with such a plasma as they propagate away from the star are strongly modified by these effects.

\subsection{Dynamical response and magnetar oscillations}

\citet{Duncan98} was the first to predict that the most energetic magnetar bursts might excite global seismic oscillations of the neutron star.   He suggested that torsional shear oscillations of the neutron star crust, with a frequency of a few tens of hertz, might be the easiest to excite and observe.  The subsequent discovery of quasi-periodic oscillations (QPOs) with the predicted frequencies during a rare and highly energetic giant flare from the magnetar SGR 1806-20 appeared to confirm this prediction in dramatic fashion \citep{Israel05}.   QPOs with similar frequencies were then found in the lightcurve of earlier giant flare from the magnetar SGR 1900+14 \citep{Strohmayer05}.    It was swiftly realised that if the frequencies of the QPOs could be identified with global seismic vibrations of neutron stars then they could be used to constrain both the composition and the internal magnetic field (see \citealt{Watts11} for a recent review).  

Since this time there has been major theoretical effort to develop models that adequately capture the complex physics that determines the vibrational frequencies of magnetars.  The picture is certainly more complicated than the initial predictions suggested, although global seismic oscillations remain the best model to explain the QPOs.    One major issue is the effect of the strong magnetic field, which acts to couple together the crust and the core, giving rise to a spectrum of magneto-elastic oscillation frequencies including both continua and discrete modes.  At present there is still disagreement between the various theoretical groups about the nature and effects of the continua on the resulting frequencies and their longevity (see for example \citealt{vanHoven11}, \citealt{Gabler11}, and \citealt{Colaiuda12}).   Uncertainties in the composition of the neutron star crust, and the role of superfluidity, will also affect frequencies \citep{Watts07, vanHoven08, Steiner09, Andersson09}.  Only once these issues have been resolved will it be possible to use the QPOs to constrain dense matter and the strong magnetic field.    However the rate of progress in terms of model development is encouraging.   

\subsection{Strong field radiation processes}

The strong magnetic field of a magnetar has a major effect on emission and radiation processes in bursts:  their temporal and spectral properties should therefore contain the signatures of high field processes.    One area in particular where the field is thought to play a major role is in setting the durations of the bursts.  While the trigger mechanism should set the rise time of the bursts, their durations are typically longer.  Does the initial trigger recur, with one event sparking the next - an
avalanche of reconnection, or a network of propagating crust ruptures? Or is there some way of
storing released energy and emitting it on a longer timescale? 

Seismic vibrations, as discussed in the previous section, are one way of
storing energy for slower release to the magnetosphere. Another well-discussed alternative is the
formation, after rapid energy release, of a plasma fireball trapped within closed field lines \citep{Thompson95}. The fireball gradually leaks energy as jets of radiation that are collimated by strong field scattering processes: this is thought
to explain the prolonged emission and rotationally modulated pulses seen in the giant flares \citep{Feroci01}.  Whether fireballs explain
the smaller bursts, however, is not known \citep{Gogus01}.  Indeed, more than one trigger and emission
mechanism may be necessary to explain burst diversity \citep{Gavriil04, Woods05}.   Recent studies of burst properties have revealed a complex picture:  simple phenomenological fits to burst spectra, for example, suggest that the emitting areas involved can be very small, and reveal interesting changes in properties over the course of outbursts \citep{Lin11, vanderHorst12, vonKienlin12}.  However these spectral models do not take into account all of the physics that we know to be relevant at high field strengths, so cannot yet be used to diagnose strong magnetic field properties \citep{Baring95, Lyubarsky02, Harding06}.   Further development of theoretical models to take advantage of the wealth of observational data now available is essential.  

Understanding the radiative environment of the star is also critical in the interpretation and analysis of the QPOs discussed in the previous section.  The observed amplitudes of the QPOs are much higher than realistic estimates for the physical amplitude of seismic vibrations of the coupled crust-core system \citep{Duncan98, Levin11}.   One possibility is that the strongly-peaked rotational pulse profile of the star plays a role \citep{Strohmayer05}.  The pulses have sharp edges, like a torch beam, so only a small physical motion may be required to move them completely in and out of the line of sight - leading to a large amplitude brightness variation.  A detailed study by \citet{Dangelo12} has now shown that although this effect will indeed enhance QPO amplitude, it is highly geometry-dependent and is unlikely to be large enough to account for all of the observed amplitudes.  This suggests the intensity itself is being modulated by some magnetospheric process, involving either the trapped fireball or subsequent scattering \citep{Timokhin08}.  Direct involvement of magnetospheric processes may solve some of the problems that are becoming apparent in the seismic models:  in particular, the difficulty of explaining the highest frequency QPOs with global magneto-elastic oscillations of a crust-core system, and the longevity of the QPOs \citep{vanHoven11, Gabler11}.  

One other issue that has attracted interest recently is the possibility of identifying the magnetic Eddington limit in magnetar bursts, either via spectral fitting \citep{Israel08, Esposito08} or via Photospheric Radius Expansion (PRE, \citealt{Watts10}).     The Eddington limit, the luminosity at which radiation pressure can balance gravity or any other confining force (such as magnetic tension), is a standard candle.  Given an independent measure of distance it can in principle be used to constrain stellar mass and radius (and hence the dense matter equation of state) and the magnetic field strength.   This technique is well-established for thermonuclear bursts from accreting neutron stars, where magnetic fields are weak \citep{vanParadijs78, Lewin93}.  For magnetar bursts derivation of the Eddington limit is more complex.  The strong magnetic field has a major effect on opacities (which are polarization dependent), and magnetic confinement may play a role \citep{Paczynski92, Ulmer94, Thompson95, Miller95}.  Nevertheless the same principle applies:  identification of the Eddington limit would provide very useful constraints on stellar properties \citep{Watts10}.

The spectral fitting arguments rely on an apparent saturation when fitting two blackbody models to magnetar burst spectra.  \citet{Israel08} argued that the two blackbodies might be interpreted as the photospheres associated with the two different polarization modes, with saturation occurring at the Eddington luminosity.    However the validity of this interpretation is not well-established \citep{Lyubarsky02, Lin11}.  The possibility of identifying PRE events, by contrast, was prompted by the observation of a pronounced double-peak structure in a bright burst from the magnetar SGR 0501+4516 (since in thermonuclear bursts PRE tends to result in a double peaked lightcurve).   In their analysis of this event, \citet{Watts10} established four basic conditions that are met during PRE in thermonuclear bursts, and explored the circumstances under which these conditions could be met in magnetar bursts, despite the very different trigger and emission mechanisms.   They concluded that PRE in magnetar bursts was indeed plausible, but that detailed studies of the stability of the emitting regions would be required to verify this.  Preliminary results from such follow-on studies indicate that atmospheric stability is problematic once luminosities reach the Eddington limit (van Putten et al. in preparation).  The consequences for observational signatures of the Eddington limit in magnetar bursts are still being explored.

\section{Conclusions}

Violent dynamical events that shake up the neutron star can be useful tools to study both the composition of the star (which depends on highly uncertain nuclear physics) and strong magnetic fields.    For magnetar bursts, however, there are many open theoretical questions motivated by a wealth of excellent observational data.  The way in which stress builds up in the system and the nature of the trigger for the bursts remains unknown.  The dynamical response of the star, which appears to include the possibility of exciting global seismic vibrations, is far more complex than originally envisaged.   However the possibility of identifying robust signatures of magnetic fields above the quantum critical limit, and of being able to use seismology to study neutron star interiors, are strong motivating factors for the determined astrophysicist.

\acknowledgements ALW acknowledges support from a Netherlands Organization for Scientific Research (NWO) Vidi Fellowship.

\bibliography{wattsbib}

\end{document}